\documentclass[12pt]{article}
\usepackage{graphicx,amsmath}
\usepackage{units}
\usepackage{color}

\newlength{\dinwidth}
\newlength{\dinmargin}
\newcommand{\dif}{\mathrm{d}}
\newcommand{\diff}[1]{\frac{\mathrm{d}#1}{#1}}

\def\lapproxeq{\lower .7ex\hbox{$\;\stackrel{\textstyle                                                    
<}{\sim}\;$}}                                                    
\def\gapproxeq{\lower .7ex\hbox{$\;\stackrel{\textstyle                                                    
>}{\sim}\;$}}                                                    
\def\be{\begin{equation}}                                                    
\def\ee{\end{equation}}                                                    
\def\bea{\begin{eqnarray}}                                                    
\def\eea{\end{eqnarray}}

\def\sh{\hat s}
\def\sh2{{\hat s}^2}

\def\Q{h}
\begin{document}

\begin{flushright}                                                    
IPPP/14/40  \\
DCPT/14/80 \\                                                    
\today \\                                                    
\end{flushright} 

\vspace*{0.5cm}

\begin{center}
{\Large \bf Evolution in opening angle combining}\\
\vspace*{0.5cm}
{\Large \bf DGLAP and BFKL logarithms}\\

\vspace*{1cm}
                                                   
E.G. de Oliveira$^a$, A.D. Martin$^b$ and M.G. Ryskin$^{b,c}$  \\                                                    
                                                   
\vspace*{0.5cm}                                                    
$^a$ Departamento de F\'{i}sica, CFM, Universidade Federal de Santa Catarina, 
C.P. 476, CEP 88.040-900, Florian\'opolis, SC, Brazil \\
$^b$ Institute for Particle Physics Phenomenology, University of Durham, Durham, DH1 3LE \\                                                   
$^c$ Petersburg Nuclear Physics Institute, NRC Kurchatov Institute, Gatchina, St.~Petersburg, 188300, Russia \\          
                                                    
\vspace*{1cm}

\begin{abstract} 

We  present an evolution equation which simultaneously sums the leading BFKL and DGLAP logarithms for the {\it integrated} gluon distribution in terms of a single variable, namely the {\it emission angle} of the gluon.
This form of evolution is appropriate for Monte Carlo simulations of events of high energy $pp$ (and $p\bar{p}$) interactions, particularly where small $x$ events are sampled.

\end{abstract}                                                        
\vspace*{0.5cm}                                                    
                                                    
\end{center}

\section{Introduction  \label{sec:1}} 
The aim is to devise an evolution equation for PDFs in the low $x$ region which simultaneously incorporates, at the same level, both the DGLAP and BFKL leading logarithms.  There has been attempts in this direction, which, however have not been very convenient  \cite{LR,Mar}. In the Gribov-Levin-Ryskin \cite{LR} paper the result was written in terms of an integral over Mellin moments and anomalous dimensions, while Marchesini \cite{Mar} attempted to improve the CCFM equation by working in terms of highly unintegrated distributions which depended on six arguments.

Procedures to combine BFKL and DGLAP effects, based on CCFM, were implemented in the `Small $x$' Monte Carlo \cite{MW1} and in the `CASCADE' Monte Carlo \cite{Jung1}. These Monte Carlos were written in terms of an `effective' transverse momentum, labelled $q'$ and $\bar{q}$ respectively, both variables being proportional to the square root of the gluon emission angle. However in \cite{MW1} the finite terms in the DGLAP gluon-gluon splitting function were neglected; and in \cite{Jung1} there was no possibility to include the {\it full} DGLAP contribution, which is included in the evolution equation proposed here.

Another  possibility to unify the BFKL and DGLAP equations was proposed by Kwiecinski et al. \cite{KMS}, where the role of the BFKL contribution was studied for the deep inelastic structure function $F_2$.  However, there, an integral equation was proposed for the unintegrated parton distribution. The equation was written in terms of the usual $x,k_t$ variables, and was not converted into the form of an evolution equation.
It was already noted by Ciafaloni \cite{Ci} that ordering in emission angle, provided by the coherence effect, plays an important role. Indeed this angular ordering was the basis of the CCFM integral equation.
However, evolution in terms of the opening angle was not discussed.

Here we start with the integral equation analogous to that in \cite{KMS}, and based on this equation, we show how it is possible to obtain an expression which describes the evolution in angle of the emitted parton with respect to the initial proton direction (in the infinite momentum frame). The momentum of the parton transverse to the direction of the proton is denoted by $k_t$.  A good feature of this evolution is that angular ordering of successive emissions is naturally provided by coherence effects. Therefore already at LO the results should be closer to experimental application.  Another point is that the angular variable, $\theta=k_t/xp$, accounts for both DGLAP and BFKL large logarithmic intervals; log $k_t$ in DGLAP and log$(1/x)$ in BFKL.
The evolution equation for PDFs is thus written, in terms of only two arguments -- the emission angle $\theta$ and the momentum fraction $x$. In this sense its form is very close to the conventional evolution equations. So it should be straightforward to implement.

In the present paper we consider only LO evolution; that is the simultaneous summation of LO BFKL and LO DGLAP logarithms. However, it should be possible to follow the same logic so as to include the known NLO BFKL and DGLAP effects.

\section{Unified BFKL-DGLAP evolution}
As mentioned above, following Ref.\cite{KMS}, we start with a `unified' BFKL-DGLAP evolution equation for the {\it unintegrated} gluon distribution, $f(x,k_t)$, written in integral form
\be
\label{eq:1}
f(x,k_t)=f_0(x,k_t)+\frac{\alpha_s}{2\pi}\left( \int^\infty_0 d^2k'_t\int_x^1 \frac{dx'}{x'}{\cal K}(k_t,k'_t)f(x',k'_t)+\int_{Q_0^2}^{k^2_t}\frac{dk_t^{'2}}{k_t^{'2}}\int_x^1 dz P(z)f\left( \frac{x}{z},k'_t\right)-DL \right),
\ee
where the first term on the right hand side is the input distribution, the second and third terms are the BFKL and DGLAP contributions, respectively, written using the usual DGLAP and BFKL variables. The final term, $DL$, denotes the subtraction of the double logarithmic contribution, $\int(dx'/x')(dk_t^{'2}/k_t^{'2})$, hidden in both the DGLAP and BFKL terms, which to avoid double counting needs to be subtracted.
It is best to subtract it from the BFKL part, since the DGLAP contribution already satisfies the energy-momentum sum rule. Note that the LO BFKL term produces more energy in the final state than there was in the incoming state.  So anyway we need to correct for this.
After the subtraction of the double log term $DL$, the $k_t'$ integral in the BFKL part is no longer logarithmic. The original BFKL kernel $\cal K$ is replaced symbolically by
\be
\label{eq:2}
\overline{{\cal K}}(k_t,k_t')={\cal K}(k_t,k_t')-\frac{2N_c}{k_t^{'2}}~,
\ee
where the kernel $\overline{{\cal K}}(k_t,k_t')$ acts as
\be
\label{k-bar}
\overline{{\cal K}}(k_t,k_t')f(x',k'_t)=
2N_c\frac{k^2_t}{k^{'2}_t}\left[\frac{f(x',k'_t)-f(x',k_t)}{|k^{'2}_t-k^2_t|}+
\frac{f(x',k_t)}{\sqrt{4k^{'4}_t+k^4_t}}-\frac{f(x',k'_t)}{k^2_t} \right]~.
\ee
Recall that the (LO) BFKL part of the equation sums the leading $\alpha_s$ln$(1/x)$ contributions. However, there is an important kinematical constraint. For a real emission \cite{Ci,KMS1,Bo,KMS} 
\be
k_t^{'2}<\frac{k^2_t}z  ~,~~~~~{\rm where}~~~z=x/x',
\label{eq:3}
\ee
which actually sums  
an essential part of the higher-order corrections.
 The constraint arises from the fact that, for larger values of $k'_t$, the longitudinal part
of the gluon virtuality would spoil the logarithmic structure of the integral; note that for LO BFKL we assume that the virtuality $k^2\simeq k^2_t$.
Thus the expression (\ref{k-bar}) should be rewritten as
$$\overline{{\cal K}}(k_t,k_t')f(x',k'_t)=
2N_c\frac{k^2_t}{k^{'2}_t}\left[\frac{\Theta(k^2_t/z-k^{'2}_t)f(x',k'_t)-f(x',k_t)}{|k^{'2}_t-k^2_t|}~+~\right.$$
\be
\label{eq:k-con}
\left. ~+~
\frac{f(x',k_t)}{\sqrt{4k^{'4}_t+k^4_t}}~-~\frac{\Theta(k^2_t-k^{'2}_t)f(x',k'_t)}
{k^2_t} \right]~.
\ee
 
 Note that in the last term of (\ref{eq:k-con}) we subtract the $DL$ term with the $\Theta$ function, which limits the available $k'_t$ interval, corresponding to DGLAP $k_t$ ordering.  
 After this subtraction the BFKL part does not contain the $DL$ contribution equivalent to that in the DGLAP part. 
 Incidentally, therefore, the BFKL kernel still retains a $DL$ contribution coming from the interval 
 $k_t^2 < k_t^{'2} < k_t^2/z$ which does not occur in DGLAP. In this way double counting is avoided. 
 
Strictly speaking, the BFKL kernel, $\overline{\cal K}$, depends on the azimuthal angle\footnote{For the DGLAP contribution we have a flat $\phi$ dependence from the beginning, due to strong $k_t$ ordering.} $\phi$ between $k_t$ and $k_t'$. However here, for simplicity, in order not to introduce another variable, we have already integrated over $\phi$ assuming a flat $\phi$ dependence of $f$. That is, we consider only the zero harmonic, which corresponds to the rightmost intercept~\footnote{It was demonstrated in~\cite{Motyka} that the full BFKL amplitude is well approximated by the sum of the leading `zero' harmonic contribution and simple two-reggeized-gluon exchange.}.


\section{Evolution in $\theta$}
Our aim is to obtain an evolution equation for the {\it integrated} gluon distribution, $g(x,\theta)$, which contains both BFKL and DGLAP logarithms, in terms of the single variable -- the gluon emission angle\footnote{Recall that some Monte Carlo generators actually make use of the angular variable. However, while the HERWIG Monte Carlo \cite{HERWIG} accounts for DGLAP evolution, it neglects the BFKL contribution (and the higher-twist BFKL effects), whereas the CASCADE Monte Carlo \cite{Jung1} does not include the {\it full} DGLAP splittings.} $\theta$. That is, a `unified' evolution  equation for $dg(x,\theta)/d\ln\theta$. 
The relation between the (conventional) integrated gluon distribution, $g$, and the distribution, $f$, unintegrated over its transverse momentum is
\be
xg(x,k_t^2)=\int^{k_t^2} \frac{dk_t^{'2}}{k_t^{'2}}~f(x,k_t^{'2}).
\ee
If we express this in terms of $\theta$, we have
\be
\label{eq:F}
xg(x,\theta)=\int^{\theta^2} f(x,\theta')\frac{d\theta^{'2}}{\theta^{'2}}.
\ee 
Thus we should replace $k_t$ and $k'_t$ in (\ref{eq:1}) by $\theta=k_t/xp$ and $\theta'=k_t'/x'p$.  
Now, it is convenient in the DGLAP term to replace the logarithmic integration $\int (dk_t^{'2}/k_t^{'2})$ by the logarithmic integration $2\int(d\theta'/\theta')$.
Then the DGLAP part in (\ref{eq:1}), written in terms of $(x,\theta)$ variables, has the same form as before.

 When we change the limit of integration in (\ref{eq:F}) to $\theta_1=\theta+d\theta$ we have the 
 usual DGLAP contribution, equivalent to the replacement 
 $\ln(k^2_{1t})=\ln(k_t^2)+2d\theta/\theta$, plus the contribution from the BFKL part
arising from the increase of the available $\ln(1/x')$ interval; $d\ln(1/x')=d\ln(\theta)$. Indeed, for a relatively large $k_t$, the condition $\theta' < \theta$ in (\ref{eq:F}) limits the part of the $x'$ domain in (\ref{eq:1}). 

   Note that, to LO accuracy, after the subtraction shown in (\ref{eq:2}), we may neglect the variation of $\ln(k^2_t)$ in the BFKL part, since now we do not have a logarithmic $dk_t^2/k_t^2$ integration here. For this reason we may replace $dx'/x'$ in the BFKL part of (\ref{eq:1})  by  $d\theta'/\theta'$. Hence we may write (\ref{eq:1}) for the unintegrated distribution $f(x,\theta)$, in the form
\be
f(x,\theta)=f_0(x,\theta_0)+\frac{\alpha_s}{2\pi}\int^{\theta}_{\theta_0}  \left( \int_0^\infty d^2k'_t~\overline{\cal K}(k_t,k'_t)f(x',k'_t=x'p\theta')+2\int^1_{z_{\rm min}} dz P(z)f\left(\frac{x}{z},\theta'\right)   \right)\frac{d\theta'}{\theta'},
\label{eq:uPDF}
\ee 
where $\theta_0$ is the starting point of the evolution. The input function is fixed $f_0(x,\theta_0)$. Recall that actually the upper limit in $k'_t$ integral for the real gluon emission is fixed by the $\Theta$-functions in (\ref{eq:k-con}). Since $k_t=xp\theta$ and $k'_t=x'p\theta'$, the value of the argument $x'$ in the BFKL part is 
$x'=k'_t/(p\theta')$, and correspondingly $z=x/x'=k_t\theta'/k'_t\theta$~\footnote{The condition $z<1$ means that for low $k'_t<k_t$, the upper limit of $\theta'$ in the BFKL part is not $\theta$, but is $\theta'_{\rm max}=\theta k'_t/k_t$.}.   The lower limit of the $z$ integration in the DGLAP part is given by
\be
z_{\rm min}~=~{\rm max}~(\theta'/\theta,~x),
\label{eq:zmin}
\ee
which on one hand provides the correct $k'_t=x'p\theta' <k_t=xp\theta$ DGLAP ordering, while on the other hand, ensures that the longitudinal momentum fraction $y=x/z<1$.  Now, we discuss the limits of the $k'_t$ integration in the BFKL part.  For the real gluon emission term the upper limit is prescribed by the first $\Theta$ function in (\ref{eq:k-con}), but runs up to infinity in the virtual loop correction which reflects gluon Reggeisation.  Note that these integrals are convergent. We may put the lower limit of the $k'_t$ integration as $k_0$ in order not to enter the non-perturbative domain. However, with reasonable extrapolation of the gluon density into the region $k'_t<k_0$ (as described by (\ref{eq:ex1}) or (\ref{eq:ex2}) below), the integral may, in fact, be extended down to $k'_t=0$. 

Since the LO contribution is now written in terms of an integral over $d\theta'/\theta'$, it appears that we may be able to find an  evolution equation in the usual derivative form for the {\it integrated} distribution $g(x,\theta)$. That is, it seems that we may be able to obtain an evolution equation for $dg(x,\theta)/ d{\rm ln}\theta^2$.  But, first, we have some points we must investigate. 

\subsection{Ensuring the evolution is for an integrated distribution}
Usually the evolution equation is written completely in terms of the integrated parton distributions.  For example
\be
\frac{\partial ~{\rm PDF}(x,Q^2)}{\partial \ln Q^2}~=~\frac{\alpha_s}{2\pi}\int^1_x dz ~P(z)~{\rm PDF}\left(
\frac xz,Q^2\right)\ .
\ee
On the contrary, in (\ref{eq:uPDF}) we deal with unintegrated gluon densities, as was convenient for the BFKL equation.
As a result, the value of derivative over $\ln\theta^2$, that is the unintegrated distribution in the left-hand side of (\ref{eq:uPDF}), is calculated using
not only the PDFs at the same $\theta$ angle (or $k_t$), but involves distributions at other angles $\theta'$. This is a common property of the BFKL equation (see (\ref{eq:1}), where the right-hand side contains an integration over $k'_t$). 

Actually, this is not a problem, since the unintegrated distribution which enters (\ref{eq:uPDF}) is measured at values of $\theta'<\theta$ where the derivative, $\partial ~{\rm PDF}(x,\theta)/\partial \ln\theta$, is already known from the previous evolution starting from a very small $\theta=\theta_0$.  If we start the evolution from a small value of $k_t$ (that is, a small angle $\theta$), then at each step of the evolution with a larger $\theta$ we will already know the distributions corresponding to lower values $\theta' < \theta$. But we still have to check that only smaller values of $\theta' < \theta$ enter (\ref{eq:uPDF}). Indeed, in the DGLAP part we have $\theta'=z\theta < \theta$. Moreover, in BFKL part we have the kinematical constraint, $k^{'2}_t<k^2_t/z$ of (\ref{eq:3}), which gives
\be
\theta'=\frac{ z k'_t}{xp} ~<~ \sqrt z \frac{k_t}{xp} ~<~\sqrt z \theta.
\ee
 Strictly speaking, this constraint is valid only for real emissions. On the 
other hand, in the virtual part (which describes gluon reggeization) the unintegrated distribution on the right-hand is taken at the same $k_t$ point as that on the left-hand side of the BFKL equation.  So, again, we never face values of $\theta' > \theta$.

This is an advantage of the evolution in terms of $\theta$ in comparison with the conventional evolution in terms $k_t$ (or $k^2$). In the latter ($k_t$) case, we face a contribution from $k'_t>k_t$ in the BFKL part.
\footnote{
An alternative way to see that the evolution in $\theta$ can be written in terms of integrated densities is to take the integral ``by parts'', based on the relation $d(u)v=d(uv)-ud(v)$, see \cite{OMR4}.}

Let us return to equation (\ref{eq:uPDF}). If, for the moment, we omit the quark contribution in the DGLAP part, then the equation can be written in the form
\be
\frac{\partial(xg(x,\theta))}{\partial {\rm ln}\theta^2}=
f(x,\theta)=f_0(x,\theta_0)+\frac{\alpha_s}{2\pi}\int^{\theta}_{\theta_0}  \int_0^\infty d^2k'_t~\overline{\cal K}(k_t,k'_t)f(x'=\frac xz,\theta')\frac{d\theta'}{\theta'}+\int^1_x dz P(z)\frac xz g\left(\frac{x}{z},z\theta\right)\ ,
\label{eq:theta}
\ee 
where $f_0$ accounts for the possible (infrared) contribution coming from $k'_t<k_0$, and where we already have used (\ref{eq:F}) in the final (DGLAP) term, accounting for the fact that, for a fixed longitudinal momentum fraction, $x/z$, the maximum allowed value of $\theta'$,  which satisfies DGLAP ordering $k'_t<k_t$, is $z\theta$. Since, now in the DGLAP part we have $\theta'<z\theta$, the lower limit $z_{\rm min}=x$. The argument $x'$ in the BFKL term is calculated from $k'_t$ as $x'=x(k'_t\theta/k_t\theta')$ (that is $z=k_t\theta'/k'_t\theta$); and  $f(x',\theta)=\partial[x'g(x',\theta)]/\partial\ln\theta^2$,  with $x'$ fixed according to (\ref{eq:F}).

 Recall after the subtraction (\ref{eq:2}), the integral over $k'_t$ does not 
have a logarithmic form, and is well convergent for $k'_t \ll k_t$. So, as far as we consider  sufficiently large $k_t$ (where perturbative QCD is valid), we may treat the contribution from the non-perturbative low $k'_t$ domain as  `power' corrections.  To be more precise, working at not such large $k_t$ one may extrapolate the unintegrated gluon for $k_t<k_0$ using
\be
f(x,k_t<k_0)=\frac{k^2_t}{k^2_t+k^2_a}\frac{k^2_0+k^2_a}{k^2_0}f(x,k_0),
\label{eq:ex1}
\ee
or 
the extrapolation in terms of integrated gluons
\be
xg(x,k_t<k_0)=\frac{k^2_t}{k^2_t+k^2_a}\frac{k^2_0+k^2_a}{k^2_0}xg(x,k_0),
\label{eq:ex2}
\ee
where $k_a$ is a parameter (see also~\cite{KMS}). The parameter $k_a$ (or even $k_a(x)$) may be used to provide a better matching between the derivative of $f$ at a small $k_t<k_0$ and that generated by the evolution equation in $k_t>k_0$ domain. Recall that confinement will nullify any coloured contribution, and correspondingly any parton distribution, at large distances, that is, for $k_t \to 0$.

Using the extrapolation (\ref{eq:ex1},\ref{eq:ex2}), one may perform a new global parton analysis. For input we need to parametrize the DGLAP-like parton distribution at $k_t=k_0$ {\em only} in some limited interval of $1>x>x_0$. Then the DGLAP part of the evolution will provide the input for the BFKL part at $x=x_0$ at all $k_t>k_0$, while the contribution for $k_t<k_0$ will be given, say, by 
(\ref{eq:ex1}). Now all the energy- (i.e. $1/x$-) dependence at small $x<x_0$  will be driven by the BFKL part of the equation, and not by the input distribution as in conventional DGLAP evolution.

Finally, we should mention that since the infrared domain is limited by the value of $k_t<k_0$, and not defined in terms of the angle $\theta$, the evolution (\ref{eq:theta}) should be considered only in the region of $\theta>k_0/xp$, and not at some $\theta>\theta_0$ domain with $\theta_0=const$. Of course, formally, in infinite momentum frame the initial momentum $p\to \infty$; so any $\theta_0=const$ is acceptable. Nevertheless, it would be better to bear in mind the realistic condition $\theta>k_0/xp$.\footnote{At first sight, it appears that working in terms of $\theta$ we get a result which depends explicitly on the incoming proton momentum $p$. This is not completely true. For a very large
$p$ the logarithm of angle ($\ln\theta$) plays the role of (pseudo)rapidity,
 and under variation of $p$ the argument $\ln\theta^2$ is simply shifted by a
constant value.}

\subsection{Energy-momentum conservation}  
While the DGLAP evolution conserves the energy (and the flavour) of system of partons this is not true for the LO BFKL equation. Formally in the leading $\ln(1/x)$ approximation an additional energy of the new partons is negligibly small ($\sim 1/\ln(1/x)$), but numerically this may be not negligible effect.

 In order to provide energy-momentum conservation we may add to the LO BFKL
contribution the non-logarithmic term (analogous to the $1/\omega \to 1/\omega -1$ replacement proposed to achieve the same goal in~\cite{SCH,EL,EHW}). That is, we replace in (\ref{eq:theta}) the usual BFKL integral
\be
\frac{\alpha_s}{2\pi}\int_x^1 \frac{dx'}{x'} \int_0^\infty 
 d^2k'_t~\overline{\cal K}(k_t,k'_t)f(x'=\frac xz,k'_t) 
\ee
by
\be
\frac{\alpha_s}{2\pi}\left(\int_x^1 \frac{dz}z 
 \int_0^\infty d^2k'_t~\overline{\cal K}
(k_t,k'_t)f(\frac xz,k'_t) -   \int_0^1dz\int_0^\infty d^2k'_t~\overline{\cal K}
(k_t,k'_t)f(x,k'_t)\right) .
\label{eq:sustr}
\ee
Unfortunately, in (\ref{eq:sustr}), we cannot replace the second integral by 1  ($\int_0^1dz=1$) since we have to account for the kinematical limit (\ref{eq:3}) in the part of the BFKL kernel corresponding to real emission. Therefore the integral over $z$ is written explicitly.

 A problem is that in the second term of (\ref{eq:sustr}) we now sample the 
 region  $\theta'>\theta$, since the function $f(x,k'_t)$ depends
 on $x$ and not on $x/z$. Recall, however, that after the subtraction of the 
leading double-logarithmic term (which was included in the DGLAP part) the violation of energy conservation  in the remaining BFKL part is rather small, and is caused only by {\em next-to-leading corrections}. Thus formally, at LO level, we may neglect the second term of (\ref{eq:sustr}); that is, the term which restores  energy conservation. However, since the integral over $k'_t$ is well convergent for $k'_t>k_t$, it is sufficient in the second term of (\ref{eq:sustr}), just to take a simple extrapolation into the $\theta'>\theta$ domain using, at each value of $x$,  the `frozen' 
 anomalous dimension of the unintegrated gluon density, $f(x,k'_t)$.   
To be more precise, we may in fact ensure exact energy-momentum conservation by  performing a few iterations; where the previous iteration provides the values of $f(x,k'_t)$ for $\theta'>\theta$.
 

Thus, finally, the $\theta$-evolution of the `integrated' gluon distribution has the form
 \bea
\nonumber
\frac{\partial[xg(x,\theta)]}{\partial {\rm ln}\theta^2}&=&f_0(x,\theta_0)\\
\nonumber
&+&\frac{\alpha_s}{2\pi}\left[\int^{\theta}_{\theta_0}  
\int_0^\infty d^2k'_t~\overline{\cal K}(k_t,k'_t)\frac{\partial[x'g(x',\theta')]}{\partial\ln\theta^{'2}}
\frac{d\theta'}{\theta'}-  \int_0^1dz\int_0^\infty d^2k'_t~\overline{\cal K}(k_t,k'_t)\frac{\partial[xg(x,\theta')]}{\partial\ln\theta^{'2}} \right.\\
&+&\left. \int^1_x dz P(z)\frac xz g\left(\frac{x}{z},z\theta\right) \right]\  ,
\label{eq:theta-f}
\eea
 where $x'=k'_t/p\theta'$ in the first term in [...] and $\theta'=k'_t/xp$ in 
the second term. According to (\ref{eq:F}) the derivatives 
 $\partial[x'g(x',\theta')]/\partial\ln\theta'$ (or 
 $\partial[xg(x,\theta')]/\partial\ln\theta'$ in the second term) are taken 
at fixed $x'$ (or $x$). The limit $z_{\rm min}$ is given in (\ref{eq:zmin}).

\begin{figure} 
\begin{center}
\includegraphics[height=10cm]{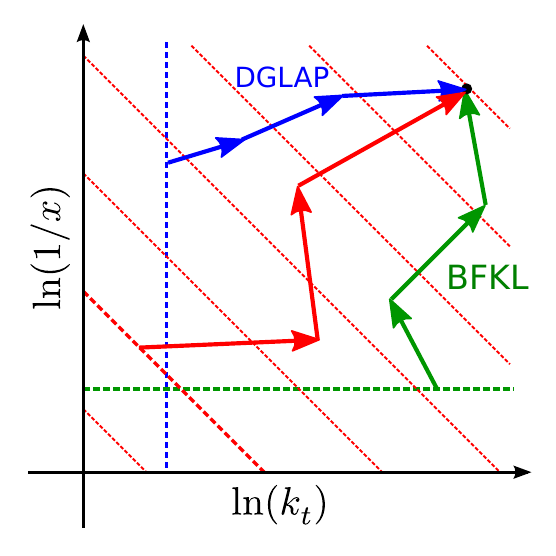}
\caption{\sf Evolution in $\theta$ unifying DGLAP and BFKL.  Each diagonal dashed line corresponds to a different fixed value of $\theta$, with the value of $\theta$ increasing towards the upper-right corner of the plot.
The upper near-horizontal path is an example of DGLAP evolution, where $k_t$ gets successively larger $k_t\gg k'_t...$, but $x$ gets a bit smaller, $x\lapproxeq x'$. Similarly the near-vertical path is an example of BFKL evolution where $x$ gets successively smaller $x\ll x'...$ with random walk in $k_t$. Unified evolution subsumes all paths with $\theta>\theta'$ to reach the point $(x,\theta)$, such as the central path shown.}
\label{fig:1}
\end{center}
\end{figure}

For illustration, in Fig. \ref{fig:1} we sketch possible evolution paths in the ln$k_t-$ln$(1/x)$ plane.  The three paths shown are examples of pure DGLAP evolution, pure BFKL evolution and unified evolution in $\theta$.

Notice from Fig. \ref{fig:1}, that to obtain a  PDF at small $x$ using DGLAP evolution we have to start evolving from an input distribution at rather low $x$ from the beginning. Analogously, in the BFKL case, to obtain a large $k_t$ gluon PDF, we need to start evolving from large $k_t$. Of course, both DGLAP and BFKL contain the double log terms which allow DGLAP to evolve from large $x$ (and BFKL to evolve from low $k_t$). However, for example in the DGLAP case, if we start from large $x$, then we will generate a PDF
$\propto \exp(\sqrt{(4\alpha_s N_c/\pi)\ln(1/x) \ln Q^2})$, 
but never containing a power of $x$, that is,  never\footnote{We could put $x^{-\lambda}$ in the input distribution, but then $\lambda$ is arbitrary, and not generated by BFKL dynamics.} one of the form $x^{-\lambda}$.   The evolution in $\theta$ will be more physical, since it starts from a region of relatively large $x$ and low $k_t$. This is more natural for an input PDF,  which is driven by physics at large distances $(\sim 0.5$ fm), corresponding to a parton confined inside a proton.

\subsection{The Sudakov $T$-factor}
Up to now we assumed that the upper scale, $\mu$, corresponding to the `hard' matrix element is of the order of $k_t$. If in some situation we will have a much higher scale $\mu \gg k_t$, then we have to account for the Sudakov suppression. That is to multiply the result by the probability that no other partons {\it (which will change the final values of $k_t$ and $x$)} are produced during the DGLAP evolution from scale $k_t$ up to the hard scale $\mu$. This probability is given by so-called $T$-factor 
\begin{equation}
T_a(k_\theta,\mu) = \exp\left(
-\int_{k_\theta^2}^{\mu^2}\!\diff{\kappa^2}\,\frac{\alpha_S(\kappa^2)}
{2\pi}\,
\int_0^1\!\dif{\zeta }\;~\zeta \sum_b \tilde{P}_{ba}(\zeta,\Delta) \right)\ ,
\label{eq:Sud}
\end{equation}
 where $\tilde P$ denotes the part of splitting function corresponding to real emission, and 
 \be
\Delta=\frac{\kappa}{\mu+\kappa}.
\label{eq:kappa}
\ee 
Moreover,  the $1/(1-z)$ singularity in the kernel $\tilde P(z,\Delta)$
contains a function $\Theta(1-z-\Delta)$ which ensures the absence of a soft parton being emitted with opening angle larger than that, $\theta_\mu$, given by the upper scale, $\mu=\mu_F$, of the DGLAP evolution. In this way we separate the partons which occur during the evolution from those that are included in the `hard' matrix element.
Correspondingly, for the last step of the evolution, in the last (DGLAP) term of (\ref{eq:theta}) the splitting function $P(z)$ should be replaced by $\tilde P(z,\Delta)$ with $\kappa=k_t$ in (\ref{eq:kappa}); see ~\cite{KMR,WMR} for more details.




\subsection{The quark contributions}
So far we have considered just the evolution equation for the gluon parton distribution. However, the generalisation to include, besides the gluon, the evolution equations for the light and heavy quark distributions is straightforward.  These latter equations have the usual DGLAP form, with no explicit BFKL contribution. Here the BFKL effects are hidden in the incoming gluon PDF driven by the equation for the gluon. Moreover, in this form it is easy to include the heavy quark mass effects. We simply follow \cite{OMR} and obtain a full set of evolution equations, which have the symbolic form
\bea
\label{eq:b4}
\dot{g} & = & {\rm BFKL ~ term}+ P_{gg} \otimes g \: + \: \sum_{q,\bar{q}}~P_{gq} \otimes q \:
+
\: \sum_{h,\bar{h}} P_{g\Q} \otimes \Q \nonumber \\
\dot{q} & = & P_{qg} \otimes g \: + \: P_{qq} \otimes q \\
\dot{{\Q}} & = & P_{
\Q g} \otimes g \: + \: P_{\Q\Q} \otimes \Q \nonumber
\eea
where $q = u,d,s$ denotes the light quark density functions and
$\Q=c,b,t$ are the heavy-quark densities.  We have used the abbreviation
$\dot{a} = (2 \pi/\alpha_S) \partial a/\partial \ln \theta^2$.  The splitting functions involving heavy quarks are given in \cite{OMR}.

Since the splitting function corresponding to the quark to gluon transition, $P_{gq}(z)$, contains a $1/z$ singularity (analogous to that in $P_{gg}$) we have to consider a possible ``BFKL'' contribution to this $q\to g$ transition.
 Recall, however, that there is no high energy ($\ln(1/x)$) leading log BFKL 
 term for quark exchange. Therefore within our LO approximation, in 
the quark cell we have to keep only the logarithmic $dk^{'2}_t/k^{'2}_t$ (DGLAP-like) contribution with $k'_t<<k_t$. Then the only possible form of the BFKL kernel $K_{qg}(k,k')$
is again pure logarithmic $1/k^2_t$~\cite{FL}, which should be subtracted to avoid double counting. In other words, at LO level, the whole $q\to g$ splitting is completely described by the usual DGLAP term.

\section{Discussion}

It is relevant to mention how the present approach compares with that of Refs. \cite{ABF} and \cite{CCS,CCSS} and the references therein. In Ref. \cite{ABF} a small $x$ resummation of the BFKL contributions was performed for the DGLAP splitting functions, that is for the anomalous dimension.
However, the small-$x$ power behaviour is still controlled by the input distribution, and not generated by the BFKL part of the evolution. Recall that the BFKL effects go beyond the anomalous dimension, and involve higher-twist effects.
In Refs. \cite{CCS,CCSS} the DGLAP-induced contributions were resummed to obtain the correction to the BFKL-Pomeron intercept. This achieved stability of the (next-to-leading-order) BFKL intercept by resumming a major part of the higher-order contributions. The procedure is very recursive equation, taking contributions from a large region of the phase space. The improved BFKL equation was not written in terms of the evolution of integrated parton densities. Again, it was claimed that the small $x$ power behaviour is mainly controlled by the input distribution. In both approaches it was not shown that the angle is a good variable, which brings
uniformity to the different contributions to the equation.

Our aim is different. We wish to determine an evolution equation for an {\it integrated} gluon distribution, which simultaneously sums both the leading BFKL and DGLAP logarithms, in terms of a {\it single} variable. We have shown that the appropriate variable is the emission angle, $\theta$, of the emitted gluon; giving an evolution equation for $\partial g(x,\theta)/ \partial {\rm ln}\theta^2$.  This novel equation is given by (\ref{eq:theta-f}) (or (\ref{eq:b4}), when the quark contribution is included).
 It brings
uniformity to the two different contributions to the equation.
 A crucial observation is that, although the right-hand side depends on $g(x',\theta')$, this does not pose a problem, since the contribution comes from $\theta'<\theta$ where $g(x',\theta')$ is known from the previous evolution.

Recall, that the inequality $\theta'<\theta$ is provided by the kinematical constraint $k^{'2}_t<k^2_t/z$ of (\ref{eq:3}), which simultaneously accounts for the major part of the higher-order BFKL next-to-leading contribution \cite{KMS1}. Besides this, we add to the BFKL part of our equation the next-to-leading term which provides the energy-momentum conservation.

The evolution in $\theta$ for the {\it integrated} gluon distribution, $g(x,\theta)$, is in contrast to the conventional BFKL equation, which is written for the {\it unintegrated} gluon distribution, $f(x,k_t^2)$.  In this case there is diffusion in log$k_t^2$ to larger values of $k_t$, as well as smaller $k_t$ and in terms of $k'_t$ integrals we have the contribution from $k'_t>k_t$.  Rather, $\theta$ in the natural variable for evolution of an integrated distribution. 

This form of `integrated' evolution in terms of a {\it single} variable should be convenient for implementation in Monte Carlo simulations of events for high energy $pp$ (and $p\bar{p}$) collisions, particularly where small $x$ events are sampled.
For instance, it would be useful to have the possibility to implement in a Monte Carlo generator the PDFs obtained independently from a global parton analysis,  based on the angular evolution proposed here. Instead, for example, the gluon PDF used by CASCADE \cite{Jung1} is evolved and fitted by the same CASCADE Monte Carlo description of a limited set of data.

\section*{Acknowledgements}

MGR thanks the IPPP at the University of Durham for hospitality. This work was supported by the Federal Program of the Russian State RSGSS-4801.2012.2.

\end{document}